\documentclass[english]{article}
\usepackage[T1]{fontenc}
\usepackage[latin9]{inputenc}
\usepackage{amsmath}
\usepackage{amsthm}
\usepackage{amssymb}
\usepackage{wasysym}

\makeatletter
\numberwithin{figure}{section}
\newcommand{\lyxaddress}[1]{
	\par {\raggedright #1
	\vspace{1.4em}
	\noindent\par}
}

\makeatother

\usepackage{babel}
\begin{document}
\title{\textbf{Black hole spectra from Vaz's quantum gravitational collapse}}
\author{\textbf{Christian Corda}}
\maketitle

\lyxaddress{\textbf{SUNY Polytechnic Institute, 13502 Utica, New York, USA, Istituto
Livi, 59100 Prato, Tuscany, Italy and International Institute for
Applicable Mathematics and Information Sciences, B. M. Birla Science
Centre, Adarshnagar, Hyderabad 500063 (India). E-mail: }\textbf{\emph{cordac.galilei@gmail.com}}}
\begin{abstract}
In 2014, in a famous paper Hawking strongly criticized the firewall
paradox by claiming that it violates the equivalence principle and
breaks the CPT invariance of quantum gravity. He proposed that the
final result of the gravitational collapse should not be an event
horizon, but an apparent horizon instead. On the other hand, Hawking
did not give a mechanism for how this could work. In the same year,
Vaz endorsed Hawking's proposal in a quantum gravitational model of
dust collapse by winning the Second Prize in the 2014 Gravity Research
Foundation Essay Competition. He indeed showed that continued collapse
to a singularity can only be obtained if one combines two independent
and entire solutions of the Wheeler-DeWitt equation. Vaz's interpretation
of the paradox was in terms of simply forbidding such a combination.
This leads naturally to matter condensing on the apparent horizon
during quantum collapse. In that way, an entirely new framework for
black holes (BHs) has emerged. The approach of Vaz was also consistent
with Einstein's idea in 1939 of the localization of the collapsing
particles within a thin spherical shell.

In this work we derive the BH mass and energy spectra via a Schrodinger-like
approach, by further supporting Vaz's conclusions that instead of
a spacetime singularity covered by an event horizon, the final result
of the gravitational collapse is an essentially quantum object, an
extremely compact ``dark star''. This ``gravitational atom'' is
held up not by any degeneracy pressure but by quantum gravity in the
same way that ordinary atoms are sustained by quantum mechanics. Finally,
by evoking the generalized uncertainty principle, the maximum value
of the density of Vaz's shell will be estimated.

\newpage{}
\end{abstract}

\section{Introduction}

Via a Gedankenexperiment, Almheiri et. al. (AMPS) \cite{key-1} claimed
that the three fundamental assumptions underlying BH Complementarity,
that are \cite{key-2,key-3}
\begin{enumerate}
\item Hawking radiation is pure;
\item Effective field theory is valid outside a stretched horizon;
\item infalling observers encounter nothing unusual as they cross the horizon;
\end{enumerate}
cannot be simultaneously consistent. In particular, AMPS proposed
that assumption 3. should be false and the infalling observer burns
up at the horizon (the horizon should act as a \textquotedblleft firewall\textquotedblright ).
Hawking strongly criticized the MPS paradox by claiming that it violates
the equivalence principle and breaks the CPT invariance of quantum
gravity \cite{key-4}. He proposed that the final result of the gravitational
collapse should not be an event horizon, but an apparent horizon instead
\cite{key-4}, but without giving a physical mechanism for how this
could work. Vaz endorsed Hawking's proposal in a quantum gravitational
model of dust collapse by winning the Second Prize in the 2014 Gravity
Research Foundation Essay Competition \cite{key-3}. In Vaz's approach,
continued collapse to a singularity can only be obtained if one combines
two independent and entire solutions of the Wheeler-DeWitt equation.
Vaz's interpretation of the paradox was in terms of simply forbidding
such a combination \cite{key-3}. The natural result was matter condensing
on the apparent horizon during quantum collapse. In that way, an entirely
new BH framework has emerged \cite{key-3}. The approach of Vaz was
also consistent with Einstein's idea in 1939 of the localization of
the collapsing particles within a thin spherical shell \cite{key-20}.

In this paper the BH mass and energy spectra via a Schrodinger-like
approach will be obtained. This further supports Vaz's conclusions
that instead of a spacetime singularity covered by an event horizon,
the final result of the gravitational collapse is an essentially quantum
object, an extremely compact ``dark star''. This ``gravitational
atom'' is held up not by any degeneracy pressure but by quantum gravity
in the same way that ordinary atoms are sustained by quantum mechanics.

\section{Schrodinger Approach to Vaz's Result}

Let us start to recall Vaz's result. Vaz realized a quantum approach
to the spherical collapse of inhomogeneous dust in AdS of dimension
$d=n+2$ \cite{key-3}, which is described by the LeMaitre-Tolman-Bondi
(LTB) family of metrics {[}5\textendash 7{]}. Via Dirac quantization
of the constraints leading to a Wheeler-DeWitt equation, Vaz found
two independent solutions in terms of shell wave functions supported
everywhere in spacetime \cite{key-3} (hereafter Planck units will
be used, i.e. $G=c=k_{B}=\hbar=\frac{1}{4\pi\epsilon_{0}}=1$)

\begin{equation}
\psi_{i}=\psi_{i}^{\left(1\right)}+A_{i}\psi_{i}^{\left(2\right)},\label{eq: wave functions}
\end{equation}
where $\psi_{i}^{\left(1\right)}$ represents dust shells condensing
to the apparent horizon on both sides of it and $\psi_{i}^{\left(2\right)}$
represents dust shells move away from the apparent horizon on either
side of it where the exterior, outgoing wave is suppressed by the
Boltzmann factor at the Hawking temperature for the shell, see \cite{key-3}
for details. One has to stress that there is nothing within the theory
that suggests a value for $A_{i}$ \cite{key-3}. In addition, further
input should be needed to determine these amplitudes \cite{key-3}.
If $0<\left|A_{i}\right|\leq1,$ then the dust will ultimately pass
through the horizon via a continued collapse arriving at a central
singularity \cite{key-3}. Consequently, an event horizon will form,
with emission of thermal radiation in the exterior \cite{key-3}.
In order to avoid the AMPS firewall paradox, $\left|A_{i}\right|$
must vanish \cite{key-3}. Then, $\psi_{i}^{\left(1\right)}$ alone
results to be the complete description of the quantum collapse \cite{key-3}.
But the meaning of $\psi_{i}^{\left(1\right)}$ is that each shell
will condense to the apparent horizon, by stopping the gravitational
collapse \cite{key-3}. Thus, there should be no tunneling into the
exterior and no AMPS firewall paradox \cite{key-3}. Each shell converges
to the apparent horizon and a \textquotedblleft dark star\textquotedblright{}
forms \cite{key-3}. 

Now, one can find the BH mass and energy spectra of this ``gravitational
atom'' via a Schrodinger-like approach. One starts to observe that,
if both the shells described by $\psi_{i}^{\left(1\right)}$converge
to the apparent horizon by forming a \textquotedblleft dark star\textquotedblright ,
by assuming absence of rotations and of dissipation during the collapse,
such a final object will be a spherical symmetric shell. In that case,
a \textquotedblleft dark star\textquotedblright{} having mass $M$
will be subjected to the classical potential 
\begin{equation}
V=-\frac{M^{2}}{2R},\label{eq: shell potential}
\end{equation}
which is indeed the self-interaction gravitational potential of a
spherical massive shell, where $R$ is its radius \cite{key-8}. In
the current case, $R$ is nothing else than the gravitational radius,
which, in a quantum framework, is subjected to quantum fluctuations
\cite{key-9}, due also to the potential absorption of external particles
\cite{key-10}. On the other hand, Eq. (\ref{eq: shell potential})
represents also the potential of a two-particle system composed of
two identical masses $M$ gravitationally interacting with a relative
position $2R$. Thus, the spherical shell is physically equivalent
to a two-particle system of two identical masses, but, clearly, as
the BH mass $M$ does not double, one has to consider the two identical
masses $M$ as being fictitious and representing the real physical
shell. Let us recall the general problem of a two-particle system
where the particles have different masses \cite{key-11}. This is
a 6-dimensional problem which can be splitted into two 3-dimensional
problems, that of a static or free particle, and that of a particle
in a static potential if the sole interaction which is felt by the
particles is their mutual interaction depending only on their relative
position \cite{key-11}. One denotes by $m_{1}$, $m_{2}$ the masses
of the particles, by $d_{1}$, $d_{2}$ their positions and by $\overrightarrow{p}_{1}$,
$\overrightarrow{p}_{2}$ the respective momenta. Being $\overrightarrow{d}_{1}-\overrightarrow{d}_{2}$
their relative position, the Hamiltonian of the system reads \cite{key-11}
\begin{equation}
H=\frac{p_{1}^{2}}{2m_{1}}+\frac{p_{2}^{2}}{2m_{2}}+V(\overrightarrow{d}_{1}-\overrightarrow{d}_{2}).\label{eq: Hamiltonian 2 particles}
\end{equation}
One sets \cite{key-11}: 
\begin{equation}
\begin{array}{ccccc}
m_{T}=m_{1}+m_{2}, &  & \overrightarrow{D}=\frac{m_{1}\overrightarrow{d}_{1}+m_{2}\overrightarrow{d}_{2}}{m_{1}+m_{2}}, &  & \overrightarrow{p}_{T}=\overrightarrow{p}_{1}+\overrightarrow{p}_{2},\\
\\
m=\frac{m_{1}m_{2}}{m_{1}+m_{2}} &  & \overrightarrow{d}=\overrightarrow{d}_{1}-\overrightarrow{d}_{2} &  & \overrightarrow{p}=\frac{m_{1}\overrightarrow{p}_{1}+m_{2}\overrightarrow{p}_{2}}{m_{1}+m_{2}}.
\end{array}\label{eq: sets}
\end{equation}
The change of variables of Eq. (\ref{eq: sets}) is a canonical transformation
because it conserves the Poisson brackets \cite{key-11}. According
to the change of variables of Eq. (\ref{eq: sets}), the motion of
the two particles is interpreted as being the motion of two fictitious
particles: i) the \emph{center of mass}, having position $\overrightarrow{D}$,
total mass $m_{T}$ and total momentum $\overrightarrow{p}_{T}$ and,
ii) the \emph{relative particle} (which is the particle associated
with the relative motion), having position $\overrightarrow{d}$,
mass $m,$ called r\emph{educed mass}, and momentum $\overrightarrow{p}$
\cite{key-11}. The Hamiltonian of Eq. (\ref{eq: Hamiltonian 2 particles})
considered as a function of the new variables of Eq. (\ref{eq: sets})
becomes \cite{key-11}: 
\begin{equation}
H=\frac{p_{T}^{2}}{2m_{T}}+\frac{p^{2}}{2m}+V(\overrightarrow{d}).\label{eq: Hamiltonian separated}
\end{equation}
The new variables obey the same commutation relations as if they should
represent two particles of positions $\overrightarrow{D}$ and $\overrightarrow{d}$
and momenta $\overrightarrow{p}_{T}$ and $\overrightarrow{p}$ respectively
\cite{key-11}. The Hamiltonian of Eq. (\ref{eq: Hamiltonian separated})
can be considered as being the sum of two terms \cite{key-11}: 
\begin{equation}
H_{T}=\frac{p_{T}^{2}}{2m_{T}},\label{eq: Hamiltonian 1}
\end{equation}
and 
\begin{equation}
H_{m}=\frac{p^{2}}{2m}+V(\overrightarrow{d}).\label{eq: Hamiltonian 2}
\end{equation}
The term of Eq. (\ref{eq: Hamiltonian 1}) depends only on the variables
of the center of mass, while the term of Eq. (\ref{eq: Hamiltonian 2})
depends only on the variables of the relative particle. Thus, the
Schrodinger equation in the representation $\overrightarrow{D},\:\overrightarrow{d}$
is \cite{key-11}: 
\begin{equation}
\left[\left(-\frac{1}{2m_{T}}\triangle_{D}\right)+\left(-\frac{1}{2m}\triangle_{d}+V(d)\right)\right]\Psi\left(D,\:d\right)=E\Psi\left(D,\:d\right),\label{eq: Schrodinger equation two particles}
\end{equation}
being $\triangle_{\overrightarrow{D}}$ and $\triangle_{\overrightarrow{d}}$
the Laplacians relative to the coordinates $\overrightarrow{D}$ and
$\overrightarrow{d}$ respectively. Now, one observes that the reduced
mass of the previously introduced two-particle system composed of
two identical masses $M$ is
\begin{equation}
m=\frac{M*M}{M+M}=\frac{M}{2}\label{eq: massa ridotta}
\end{equation}
In that case, by recalling that in Schwarzschild coordinates the BH
center of mass coincides with the origin of the coordinate system
and with the replacements
\begin{equation}
d\rightarrow2R,\label{eq: replacement}
\end{equation}
the Schrodinger equation (\ref{eq: Schrodinger equation two particles})
becomes 
\begin{equation}
\left(-\frac{1}{2m}\triangle_{2R}+V(2R)\right)\Psi\left(2R\right)=E\Psi\left(2R\right).\label{eq: Schrodinger equation membrana}
\end{equation}
Setting 
\begin{equation}
r\equiv\frac{R}{2},\label{eq: setting}
\end{equation}
the potential of Eq. (\ref{eq: shell potential}) becomes 
\begin{equation}
V=-\frac{m^{2}}{r},\label{eq: energia potenziale membrana}
\end{equation}
and the Schrodinger equation in the representation $\overrightarrow{D}=0,\:\overrightarrow{r}$
becomes
\begin{equation}
-\frac{1}{2m}\left(\frac{\partial^{2}\Psi}{\partial r^{2}}+\frac{2}{r}\frac{\partial\Psi}{\partial r}\right)+V\Psi=E\Psi.\label{eq: Schrodinger membrana ritrovata}
\end{equation}
The Schrodinger equation (\ref{eq: Schrodinger membrana ritrovata})
is formally identical to the traditional Schrodinger equation of the
$s$ states ($l=0$) of the hydrogen atom which obeys to the Coulombian
potential \cite{key-11} 
\begin{equation}
V(r)=-\frac{e^{2}}{r}.\label{eq: energia potenziale atomo idrogeno}
\end{equation}
In the potential of Eq. (\ref{eq: energia potenziale membrana}) the
squared electron charge $e^{2}$ is replaced by the squared reduced
mass $m^{2}.$ Thus, Eq. (\ref{eq: Schrodinger membrana ritrovata})
can be interpreted as the Schrodinger equation of a particle, the
``electron'', which interacts with a central field, the ``nucleus''.
On the other hand, this is only a mathematical artifact because the
real nature of the quantum BH is in terms of Vaz's shell. For the
bound states ($E<0$) the energy spectrum is 
\begin{equation}
E_{n}=-\frac{m^{5}}{2n^{2}}.\label{eq: spettro energia}
\end{equation}
Hence, in order to completely solve the problem, one must find the
relationship between the reduced mass and the total energy of Vaz's
shell. In order to find the relationship between the reduced mass
and the total energy one can use the Oppenheimer and Snyder (OS) gravitational
collapse \cite{key-12} rather than the LTB one. One recalls that,
in the classical framework there are considerable differences in the
geometry of evolving trapped regions for OS and LTB \cite{key-22}.
In the former the boundary of the star gets trapped much before the
singularity, while for the latter there are scenarios where trapping
occurs extremely close to the central singularity and thus allowing
outgoing null geodesic from a very dense region \cite{key-22}. The
situation changes in the quantum framework. In fact, the Author and
Collaborators recently developed two different approaches to the quantum
OS collapse {[}23\textendash 25{]}. The first one \cite{key-23,key-24}
concerns the application of a quantization procedure originally developed
in a cosmological framework by the historical collaborator of Einstein,
Nathan Rosen \cite{key-16}. The latter \cite{key-25} concerns the
use of Feynman's path integral approach \cite{key-26}. In both the
cases, remarkably, it has been obtained a Schrodinger equation consistent
with Eq. (\ref{eq: Schrodinger membrana ritrovata}). This means that
the two different gravitational collapse, i.e. the OS and the LTB,
generate the same quantum object. The physical interpretation of such
a Schrodinger equation was in terms of a two particle system {[}23\textendash 25{]}.
Instead, in the current approach we argue that such a two particle
system is a fictitious mathematical artifact which hiddens the real
physical nature of the quantum BH, which surfaces in terms of a spherical
massive shell in agreement with Vaz's approach \cite{key-3}.

In the case of the gravitational collapse the internal solution is
given by the well known Friedmann-Lemaitre-Robertson-Walker line-element,
which, by using spherical coordinates and comoving time is \cite{key-14}
\begin{equation}
ds^{2}=d\tau^{2}-a^{2}(\tau)\left(\frac{dr^{2}}{1-r^{2}}+r^{2}d\theta^{2}+r^{2}\sin^{2}\theta d\varphi^{2}\right).\label{eq: FLRW}
\end{equation}
Considering the Einstein field equation \cite{key-14}
\begin{equation}
G_{\mu\nu}=-8\pi T_{\mu\nu}\label{eq: Einstein field equation}
\end{equation}
and assuming zero pressure one gets the following relations \cite{key-15}
\begin{equation}
\begin{array}{c}
\dot{a}^{2}=\frac{8}{3}\pi a^{2}\rho-1\\
\\
\ddot{a}=-\frac{4}{3}\pi a\rho
\end{array}\label{eq: evoluzione}
\end{equation}
being $\dot{a}=\frac{da}{d\tau}$. In order to have consistency, one
obtains \cite{key-16}
\begin{equation}
\frac{d\rho}{da}=-\frac{3\rho}{a}.\label{eq: consistenza}
\end{equation}
The last equation is integrated as 
\begin{equation}
\rho=\frac{C}{a^{3}}.\label{eq: densit=0000E0}
\end{equation}
The integration constant $C$ is obtained via the initial conditions
as \cite{key-14} 
\begin{equation}
C=\frac{3a_{0}}{8\pi}.\label{eq: C}
\end{equation}
Eq. (\ref{eq: densit=0000E0}) can be rewritten as 
\begin{equation}
\rho=\frac{3a_{0}}{8\pi a^{3}}.\label{eq: densit=0000E0 2}
\end{equation}
 One also recalls that the standard Einstein - Hilbert Lagrangian
is \cite{key-14} 
\begin{equation}
L_{EH}=\frac{\sqrt{-g}R}{16\pi}+L_{M},\label{eq: EH}
\end{equation}
where, as usual, $L_{M}$ represents the matter fields. For the collapsing
dust it is $L_{M}=\rho$ \cite{key-21}. If one uses the Friedmann-Lemaitre-Robertson-Walker
line-element (\ref{eq: FLRW}) one gets
\begin{equation}
L_{FLRW}=\dot{a}^{2}+\frac{8}{3}\pi a^{2}\rho.\label{eq: Lag FLRW}
\end{equation}
The Lagrangian (\ref{eq: Lag FLRW}) can be rescaled as 
\begin{equation}
L=\frac{M\dot{a}^{2}}{2}+\frac{4}{3}M\pi a^{2}\rho,\label{eq: Lag rescaled}
\end{equation}
which can be put as
\begin{equation}
L=\frac{M\dot{a}^{2}}{2}-V(a),\label{eq: Lagrangiana F}
\end{equation}
where 
\begin{equation}
V(a)\equiv-\frac{4}{3}M\pi a^{2}\rho.\label{eq: potenziale F}
\end{equation}
If one inserts Eq. (\ref{eq: densit=0000E0 2}) into Eq. (\ref{eq: potenziale F})
one gets 
\begin{equation}
V(a)=-\frac{Ma_{0}}{2a}.\label{eq: energia potenziale 2}
\end{equation}
One also finds the energy function associated to the Lagrangian as
\begin{equation}
E=\frac{\partial L}{\partial\dot{a}}\dot{a}-L.\label{eq: en lagr}
\end{equation}
Now, if one inserts Eq. (\ref{eq: Lag rescaled}) in Eq. (\ref{eq: en lagr})
and uses the first of Eqs. (\ref{eq: evoluzione}) one gets 

\begin{equation}
E=-\frac{M}{2}.\label{eq: energia totale F}
\end{equation}
A final correction is needed. Let us clarify the reason. One compares
Eq. (\ref{eq: energia potenziale membrana}) with the analogous potential
energy of the hydrogen atom, i.e. Eq. (\ref{eq: energia potenziale atomo idrogeno}).
Eqs. (\ref{eq: energia potenziale membrana}) and (\ref{eq: energia potenziale atomo idrogeno})
are almost formally identical. In fact, one recognizes a fundamental
difference. In Eq. (\ref{eq: energia potenziale atomo idrogeno})
the charge of the electron is constant for all the hydrogen atom's
energy levels. Instead, in Eq. (\ref{eq: energia potenziale membrana})
the BH mass changes during the jumps from an energy level to another
because of the emissions of quanta and the absorptions of external
particles. The BH mass indeed decreases for emissions and increases
for absorptions. Hence, one needs also to consider such a BH dynamical
behavior. A good way to realize this is via the introduction of the
\emph{BH effective mass $M_{E}$ }\cite{key-10}, which is the average
of the BH initial and final masses which are involved in a quantum
transition. \emph{$M_{E}$} indeed represents the BH mass\emph{ during}
the BH expansion (contraction), which is triggered by an absorption
(emission) of a particle \cite{key-10}. The rigorous definition of
the BH\emph{ }effective mass is \cite{key-10}
\begin{equation}
M_{E}\equiv M\pm\frac{\omega}{2},\label{eq: effective quantities absorption}
\end{equation}
being $\omega$ the mass-energy of the absorbed (emitted) particle
(the sign plus concerns absorptions, the sign minus concerns emissions).
Hence, one sees that introducing the BH effective mass in the BH dynamical
framework is very intuitive. But, for the sake of mathematical rigor,
that introduction has been completely justified via Hawking's periodicity
argument \cite{key-10}. One chooses the positive sign in Eq. (\ref{eq: effective quantities absorption})
if one considers the BH formation in terms of absorptions of external
particles in a quantum framework. Therefore, Eq. (\ref{eq: effective quantities absorption})
reads 
\begin{equation}
M_{E}\equiv M+\frac{\omega}{2}.\label{eq: effective mass absorption}
\end{equation}
In a quantum framework one wants to obtain the energy eigenvalues
as being absorptions starting from the BH formation, that is from
the BH having null mass, where with ``the BH having null mass''
one means the situation of the gravitational collapse before the formation
of the first apparent horizon. This implies that one must replace
$M\rightarrow0$ and $\omega\rightarrow M$ in Eq. (\ref{eq: effective mass absorption}).
Thus, one gets
\begin{equation}
M_{E}=\frac{M}{2}=m,\label{eq: effective quantities absorption finali}
\end{equation}
and Eq. (\ref{eq: energia totale F}) becomes 
\begin{equation}
E=-\frac{m}{2}.\label{eq: energia totale reale}
\end{equation}
By inserting this last equation in Eq. (\ref{eq: spettro energia}),
a bit of algebra permits to obtain the energy spectrum 
\begin{equation}
E_{n}=-\frac{1}{2}\sqrt{n},\label{eq: BH energy levels finale.}
\end{equation}
and the corresponding mass spectrum 
\begin{equation}
M_{n}=2\sqrt{n}.\label{eq: spettro massa BH finale}
\end{equation}

\section{Density of Vaz's Schell}

It is also important to estimate the maximum value of the density
of Vaz's shell. By considering the shell's mathematical description
of Eqs. (\ref{eq: energia potenziale membrana}) and (\ref{eq: Schrodinger membrana ritrovata})
in terms of a quantum system composed by a fictitious particle, the
``electron'', which interacts through a quantum gravitational interaction
with a central field, the ``nucleus'', the Born rule \cite{key-17}
and the Copenhagen interpretation of quantum mechanics \cite{key-18}
imply that the position of the ``electron'' cannot be exactly localized,
but it is only possible to get the probability density of finding
the ``electron'' at a given point which is, in turn, proportional
to the square of the magnitude of the wavefunction of the ``electron''
at that point. Being the system ``electron-nucleus'' only a fictitious
representation of the physical quantum shell, this implies that one
cannot exactly localize the position of the quantum shell via the
oscillating gravitational radius, and must, in turn, use an average
radius. The average radius of Vaz's shell is given by the shell's
expected radial distance \cite{key-11} 
\begin{equation}
\bar{R}_{n}=\frac{3}{2}M_{n}=3\sqrt{n}.\label{eq: size shell from mass}
\end{equation}
If one evokes the generalized uncertainty principle \cite{key-19},
which guarantees that the shell must have a physical thickness, at
least of the order of the Planck length, one can compute the minimum
volume of Vaz's shell (in Planck units) as the difference between
the volume of the sphere having radius $3\sqrt{n}+\frac{1}{2}$ and
the volume of the sphere having radius $3\sqrt{n}-\frac{1}{2}.$ Thus,
one gets: 
\begin{equation}
\begin{array}{c}
V_{min}=\frac{4}{3}\pi\left[\left(3\sqrt{n}+\frac{1}{2}\right)^{3}-\left(3\sqrt{n}-\frac{1}{2}\right)^{3}\right]\\
\\
=\frac{4}{3}\pi\left(27n+\frac{1}{4}\right)=36\pi n+\frac{\pi}{3}.
\end{array}\label{eq: volume guscio}
\end{equation}
On the other hand, the mass spectrum of the shell is given by Eq.
(\ref{eq: spettro massa BH finale}). Hence, one obtains the maximum
value of the density of the quantum shell as 
\begin{equation}
\rho_{max}=\frac{2\sqrt{n}}{36\pi n+\frac{\pi}{3}}.\label{eq: densit=0000E0 massima}
\end{equation}
The maximum density decreases with increasing $n,$ as one intuitively
expects. Thus, the maximum density corresponds to the ground state
of Vaz's shell, that, for $n=1,$ is a density of 
\begin{equation}
\rho_{max}(n=1)=\frac{2}{36\pi+\frac{\pi}{3}}\simeq0.0175,\label{eq: ground state density}
\end{equation}
in Planck units. By recalling that the Planck density is roughly $10^{93}$
grams per cubic centimetre in standard units, one gets a value of
\begin{equation}
\rho_{max}(n=1)\simeq1.752*10^{91}\:grams\:per\:cubic\:centimetre\label{eq: densit=0000E0 rozza}
\end{equation}
for the density of the ground state of Vaz's shell in standard units,
which is very high but about two order of magnitude less than the
Planck density. For large $n$ Eq. (\ref{eq: densit=0000E0 massima})
is well approximated by 
\begin{equation}
\rho_{max}\simeq\frac{1}{18\pi\sqrt{n}}.\label{eq: densit=0000E0 massima approssimata}
\end{equation}
For a BH having mass of the order of 10 solar masses Eq. (\ref{eq: spettro massa BH finale})
gives 
\begin{equation}
\sqrt{n}=\frac{10M_{\astrosun}}{2}=5M_{\astrosun}\sim\frac{10^{34}\:grams}{M_{p}}\sim5*10^{38}.\label{eq: quantum levels 10 solar masses}
\end{equation}
being $M_{\astrosun}\sim2*10^{33}\:grams$ the solar mass and $M_{p}\sim2*10^{-5}\:grams$
the Planck mass. By inserting the result of Eq. (\ref{eq: quantum levels 10 solar masses})
in Eq. (\ref{eq: densit=0000E0 massima approssimata}) one gets 
\begin{equation}
\rho_{max}(10M_{\astrosun})\sim\frac{1}{5*18\pi*10^{38}}\sim3.5*10^{-41}\label{eq: densita 10 masse solari}
\end{equation}
in Planck units and, being the Planck density roughly $10^{93}$ grams
per cubic centimetre in standard units, one finds a value of 
\begin{equation}
\rho_{max}(10M_{\astrosun})\sim3.5*10^{52}\:grams\:per\:cubic\:centimetre.\label{eq: densita 10 masse solari SU}
\end{equation}

\section{Conclusion}

Summarizing, in this work the mass and energy spectra of Vaz's quantum
shell have been obtained via a Schrodinger-like approach, by further
supporting Vaz's conclusions that instead of a spacetime singularity
covered by an event horizon, the final result of the gravitational
collapse is an essentially quantum object, an extremely compact ``dark
star''. This ``gravitational atom'' is held up not by any degeneracy
pressure but by quantum gravity in the same way that ordinary atoms
are sustained by quantum mechanics. By evoking the generalized uncertainty
principle, the maximum value of the density of Vaz's shell has been
estimated.

\section*{Acknowledgements}

The Author thanks an unknown Referee for very useful comments and
suggestions.

\section*{Conflict of Interest}

The author has declared no conflict of interest.

\section*{Keywords}

Black holes, quantum spectrum, gravitational collapse.


\begin{thebibliography}{10}
\bibitem{key-1}A. Almheiri, D. Marolf, J. Polchinski, and J. Sully,
J. High Energ. Phys. \textbf{2013}, 62 (2013).

\bibitem{key-2}L. Susskind, L. Thorlacius and J. Uglum, Phys. Rev.
D \textbf{48} 3743 (1993).

\bibitem{key-3}C. Vaz, Int. J. Mod. Phys. D \textbf{23}, 1441002
(2014). 

\bibitem[4]{key-4}S. W. Hawking, arXiv:1401.5761 (2014). 

\bibitem[5]{key-5}G. LeMa\^{\i}tre, Ann. Soc. Sci. Bruxelles I, \textbf{A53,}
51 (1933).

\bibitem[6]{key-6}R.C. Tolman, Proc. Natl. Acad. Sci., USA \textbf{20,}
410 (1934). 

\bibitem[7]{key-7}H. Bondi, Mon. Not. Astron. Soc. \textbf{107,}
343 (1947).

\bibitem[8]{key-8}R. Arnowitt, S. Deser, and C. W. Misner, Phys.
Rev. \textbf{120}, 313 (1960). 

\bibitem[9]{key-9}J. D. Bekenstein, in Prodeedings of th Eight Marcel
Grossmann Meeting, T. Piran and R. Ruffini, eds., pp. 92-111 (World
Scientific Singapore 1999). 

\bibitem[10]{key-10}C. Corda, Class. Quantum Grav. \textbf{32,} 195007
(2015).

\bibitem[11]{key-11}A. Messiah, \emph{Quantum Mechanics, Vol. 1},
North-Holland, Amsterdam (1961).

\bibitem[12]{key-12}J. R. Oppenheimer and H. Snyder, Phys. Rev. \textbf{56},
455 (1939).

\bibitem[13]{key-13}D. L. Beckerdoff and C. W. Misner, D. L. Beckerdoff\textquoteright s
A. B. Senior Thesis, Princeton Univeristy (1962).

\bibitem[14]{key-14}C. W. Misner, K. S. Thorne and J. A. Wheeler,
\emph{Gravitation} (W. H. Feeman and Co., 1973).

\bibitem[15]{key-15}S. Weinberg, \emph{Gravitation and Cosmology},
Wiley, New York (1972). 

\bibitem[16]{key-16}N. Rosen, Int. Journ. Theor. Phys. \textbf{32},
1435, (1993).

\bibitem[17]{key-17}M. Born, Zeit. Phys. \textbf{37}, 863 (1926).

\bibitem[18]{key-18}N. Bohr, Nature \textbf{121}, 580 (1928).

\bibitem[19]{key-19}M. Maggiore, Phys. Rev. D \textbf{49}, 2918 (1994).

\bibitem[20]{key-20}A. Einstein, Ann. Math. (Second Series) \textbf{40},
922 (1939). 

\bibitem[21]{key-21}S. Mendoza and S.­ Silva, Int. Jour. Geom. Meth.
Mod. Phys. \textbf{18}, 2150059 (2021).

\bibitem[22]{key-22}Private communication with the Referee.

\bibitem[23]{key-23}C. Corda and F. Feleppa, Adv. Theor. Math. Phys.
(2023), pre-print in arXiv:1912.06478.

\bibitem[24]{key-24}C. Corda, F. Feleppa and F. Tamburini, EPL \textbf{132},
30001 (2020).

\bibitem[25]{key-25}C. Corda, F. Feleppa, F. Tamburini, I. Licata,
Theor. Math. Phys. \textbf{213}, 1632 (2022).

\bibitem[26]{key-26}R. Feynman, ``Feynman's Thesis \textendash{}
A New Approach to Quantum Theory'', Edited by Laurie M Brown, World
Scientific (2005).
\end{thebibliography}
\end{document}